\documentclass[pre,twocolumn,twoside,showpacs,superscriptaddress,tightenlines]{revtex4-1}
\usepackage{graphicx,amssymb,amsmath,epsf,bm}
\epsfclipon
\usepackage{color}
\definecolor{nred}{RGB}{224,0,0}
\definecolor{nblue}  {RGB}{28,130,185}
\definecolor{dgreen} {RGB}{78,138,21}
\definecolor{norange}{RGB}{230,120,20}

\newcommand{\be}{\begin{equation}}
\newcommand{\ee}{\end{equation}}

\begin{document}

\title{Convergence of  Non-Perturbative Approximations to the Renormalization Group}

\author{Ivan Balog}
\affiliation{Institute of Physics, Bijeni\v{c}ka cesta 46, HR-10001 Zagreb, Croatia}

\author{Hugues Chat\'e}
\affiliation{Service de Physique de l'Etat Condens\'e, CEA, CNRS Universit\'e Paris-Saclay, CEA-Saclay, 91191 Gif-sur-Yvette, France}
\affiliation{Computational Science Research Center, Beijing 100094, China}
\affiliation{Sorbonne Universit\'e, CNRS, Laboratoire de Physique Th\'eorique de la Mati\`ere Condens\'ee, LPTMC, F-75005 Paris, France}

\author{Bertrand Delamotte}
\affiliation{Sorbonne Universit\'e, CNRS, Laboratoire de Physique Th\'eorique de la Mati\`ere Condens\'ee, LPTMC, F-75005 Paris, France}

\author{Maroje Marohni\'c}
\affiliation{Visage Technologies AB, Diskettgatan 11A SE-583 35 Link\"oping, Sweden}

\author{Nicol\'as Wschebor}
\affiliation{Instituto de F\'{\i}sica, Faculdad de Ingenier\'ia, Universidad de la Rep\'ublica, 11000 Montevideo, Uruguay}

\date{\today}

\begin{abstract}
We provide analytical arguments showing that the non-perturbative approximation scheme to Wilson's renormalisation group known as the derivative expansion has a finite radius of convergence.
We also provide guidelines for choosing the regulator function at the heart of the procedure 
and propose empirical rules for selecting an optimal one, without prior knowledge of the problem at stake.
Using the Ising model in three dimensions as a testing ground and the derivative expansion at order six, we find fast convergence of critical exponents to their exact values, 
irrespective of the well-behaved regulator used, in full agreement with our general arguments. 
We hope these findings will put an end to disputes regarding this type of non-perturbative methods.
\end{abstract}

\maketitle

Wilson's renormalization group (RG) is an extraordinary
means of understanding quantum and statistical
field theories. Its perturbative implementation \cite{Wilson:1973jj,Guida:1998bx,Pelissetto02},
in particular under the form of the $\epsilon$-expansion, has
been a very efficient toolbox. The last twenty-five
years, however, have witnessed a revival of Wilson's RG because an
alternative formulation \cite{Wetterich:1992yh,Ellwanger:1993kk,Morris:1993qb,Berges:2000ew,Delamotte:2007pf} 
has allowed for new and nonperturbative
approximation schemes (which are anyway needed to solve the exact RG equation).
Using this nonperturbative approach to the RG (NPRG hereafter), 
remarkable results have been obtained on problems
that were either very difficult or fully out of reach of the perturbative approach. 

The method is versatile, allowing to treat equilibrium and non-equilibrium problems, disordered systems, 
with access to both universal and non-universal quantities. 
To list a few successes just within statistical physics, let us mention the random field Ising model (spontaneous supersymmetry breaking and the associated breaking of dimensional reduction in a nontrivial dimension) \cite{Tarjus:2004wyx,Tissier:2011zz}, 
the Kardar-Parisi-Zhang equation in dimensions larger than one (identification of the strong coupling fixed point) 
\cite{PhysRevLett.104.150601,PhysRevE.84.061128,PhysRevE.86.019904}, 
the glassy phase of crystalline membranes \cite{Coquand2017}, 
systems showing different critical exponents in their high and low temperature phases \cite{Leonard15}, the phase diagram of reaction-diffusion systems \cite{PhysRevLett.92.255703,Canet:2003yu}, etc.

Most of these results were obtained using an NPRG approximation scheme known as the derivative expansion (DE). 
In a nutshell, the underlying ideas are as follows: 
The exact NPRG equation governs the evolution of an effective action $\Gamma_k$ 
(in the field theory language the generating functional of one-particle irreducible correlation functions) with the RG momentum scale $k$. In the NPRG approach, a regulator function $R_k(q^2)$ ensures that the large  wavenumber modes (with $q^2>k^2$) are progressively integrated over while the others are frozen.  When $k=0$, all statistical fluctuations have been integrated and $\Gamma_{k=0}=\Gamma$, the Gibbs free energy of the model. The DE  consists in approximating  the functional $\Gamma_k[\phi]$, where $\phi$ represents all the fields of the problem, by its Taylor expansion in gradients of $\phi$ truncated at a finite order.

In spite of its undeniable successes, the DE ---and the NPRG in general--- has often been criticized. 
Two main points are usually raised, the (apparent) lack of a small parameter controlling its convergence 
and the arbitrariness induced by the choice of the regulator function $R_k$. Indeed, within any approximation scheme,
the end results do carry a residual influence of $R_k$.
This has been often invoked against the NPRG approach, even though the dependence on $R_k$ 
is similar to the renormalization scheme dependence in perturbation theory \footnote{In perturbation theory, a set of prescriptions, either  implicit or explicit, is necessary to parameterize the theory in terms of physical quantities. In principle, the physics should be independent of this renormalization scheme, but truncating the perturbative series at any finite order induces a dependence.}.

In this Letter, we aim to put an end to this controversy. 
We use the Ising model as a testing ground both because its relative simplicity allows us to study the sixth order of the DE and because its critical exponents are accurately known \cite{ElShowk:2012ht,El-Showk:2014dwa,Kos:2014bka}. We provide numerical evidence and analytical arguments showing that the DE not only converges, but does so rapidly. This conclusion holds beyond the Ising model and is most likely generic.
Contrary to usual perturbative approaches, we find that the DE has (i) a finite radius of convergence and (ii) a fast convergence,
even at low orders, when the anomalous dimension is small.
We also discuss the respective quality of regulators $R_k(q^2)$ and propose
empirical rules for selecting  optimal ones, without prior knowledge about the problem at stake. 

We start with a brief review of the NPRG, specialized here to the $\phi^4$ model for convenience. A one-parameter family of models indexed by a scale $k$ is defined such that only the short wavelength fluctuations, 
with wavenumbers $q=|{\bf q}|> k$, are summed over in the partition function ${\cal Z}_k$. 
The decoupling of the slow modes, $\varphi(|{\bf q}|< k)$, in ${\cal Z}_k$ is performed by adding to the original Hamiltonian $H$ a quadratic (mass-like) term which is nonvanishing only for these modes:
\begin{equation}
{\cal Z}_k[J]=\int D\varphi \exp\left[-H[\varphi]- \Delta H_k[\varphi] +\int_x J \varphi   \right]
\label{partition}
\end{equation}
where $\Delta H_k[\phi]= 1/2\int_{\bf q} R_k(q^2)\varphi({\bf q})\varphi(-{\bf q})$. 
The form of the regulator function $R_k(q^2)$ is discussed in detail below
(see Eqs. (\ref{regulator-wetterich}-\ref{regulator-exp}) for examples used here). 
The $k$-dependent Gibbs free energy $\Gamma_k[\phi]$ (with $\phi(x)=\langle\varphi(x)\rangle$) is defined
 as the (slightly modified) Legendre transform of $\log{\cal Z}_k[J]$:
\begin{equation}
    \Gamma_k[\phi]+\log{\cal Z}_k[J]=\int_x J\phi-\frac 1 2 \int_{\bf q} R_k(q^2)\phi({\bf q})\phi(-{\bf q}).
    \label{legendre}
\end{equation}
The exact RG flow equation of $\Gamma_k$ reads \cite{Wetterich:1992yh,Ellwanger:1993kk,Morris:1993qb}:
\begin{equation}
\partial_t\Gamma_k=\frac 1 2 \int_{\bf q} \partial_t R_k(q^2)\left(\Gamma_k^{(2)} +R_k \right)^{-1}\hspace{-1mm}[{\bf q},-{\bf q};\phi]
\label{wetterich}
\end{equation}
where $t=\log(k/\Lambda)$ and $\Gamma_k^{(2)}[{\bf q},-{\bf q};\phi]$ is the Fourier transform of the second functional derivative of $\Gamma_k[\phi]$. 
The DE consists in solving Eq. (\ref{wetterich}) in a restricted functional space where $\Gamma_k[\phi]$ involves a limited number of gradients of $\phi$ multiplied by ordinary functions of $\phi$. 
Zeroth order is the commonly-used local potential approximation (LPA): 
only the momentum dependences present in $H$ are kept in the correlation functions. 
For the $\varphi^4$ model,  $\Gamma_k[\phi]$ is then approximated by: 
$\int_x\left( U_k(\phi)+\tfrac{1}{2} (\nabla\phi)^2 ) \right)$: only a running potential term is retained.
At order $s=6$, the ansatz for $\Gamma_k$ involves thirteen  functions (see section I of the Supplemental Material):
\begin{eqnarray}
&\hspace{-0.5cm}  \Gamma_k[\phi]&= \int d^dx \Big[U_k(\phi) + \tfrac{1}{2} Z_k(\phi) (\partial_\mu\phi)^2 \nonumber\\
&&\!\!\!\!\!\!\! + \tfrac{1}{2} W^a_k(\phi)(\partial_\mu\partial_\nu\phi)^2+\cdots+\tfrac{1}{96}  X^h_k (\phi) \left((\partial_\mu\phi)^2\right)^3 \Big].
\label{ansatz-order6}
\end{eqnarray}
The flow of all functions is obtained by inserting the ansatz  (\ref{ansatz-order6}) 
in Eq.(\ref{wetterich}) and expanding and truncating the right hand side on the same functional subspace.
In practice, this is implemented in Fourier space. For instance, we obtain from Eq.(\ref{ansatz-order6}) that  
$Z_k(\phi)=\partial_{p^2}\Gamma_k^{(2)}(p;\phi)\vert_{{\bf p}=0}$
with $\phi$ a constant field. Thus, the flow of $Z_k(\phi)$ is given by the $p^2$ term of the flow of $\Gamma_k^{(2)}(p;\phi)$.

At criticality ---the regime of interest here---, the system is self-similar: its RG flow reaches a fixed point.
In practice, the fixed point is reachable when using dimensionless and renormalized functions denoted below by lowercase letters 
($u_k, z_k, \ldots,x_{h,k}$).
We proceed as usual \cite{Berges:2000ew} 
by rescaling fields and coordinates. Here $\tilde x= k x$, $\tilde\phi(\tilde x)=\sqrt{Z^0_k}\, k^{(2-d)/2}\,\phi(x)$. 
Functions are then rescaled according to their canonical dimension 
and renormalized by $({Z^0_k})^{n/2}$ where $n$ is the number of fields they multiply in the ansatz (\ref{ansatz-order6}). 
This leads to $Z_k(\phi) = Z^0_k\, z_k(\tilde\phi)$. 
The absolute normalization of both $Z^0_k$ and  $z_k(\tilde\phi)$ is defined only once their value is fixed at a given point. 
We use the  (re)normalization condition: $z_k(\tilde\phi_0)=1$ for a fixed value of $\tilde\phi_0$. 
The running anomalous dimension $\eta_k$ is then defined by $\eta_k= - \partial_t\log Z^0_k$. 
It becomes the anomalous dimension $\eta$ at the fixed point \cite{Berges:2000ew}.

Let us now give analytical arguments in favor of the convergence of the DE and about the rapidity of this convergence. 
We continue using the $\varphi^4$ theory here, but our results are more general.
The key remark is that the momentum expansion applied to the theory away from criticality, either in the symmetric or broken phase, is known to be convergent with a finite radius of convergence. For instance, calling $m$ the mass, that is, the inverse correlation length, the $c_n$ in
\begin{equation}
    \frac{\Gamma^{(2)}(p,m)}{\Gamma^{(2)}(0,m)}=\frac{\Gamma^{(2)}_{k=0}(p,m)}{\Gamma^{(2)}_{k=0}(0,m)}=1+\frac{p^2}{m^2}+\sum_{n=2}^{\infty} c_n \left(\frac{p^2}{m^2}\right)^n
    \label{DE-massive-case}
\end{equation}
  are universal close to criticality and behave at large $n$ as $ c_{n+1}/c_n\sim -1/9$ and $-1/4$ in the symmetric and broken phases respectively (see, for example, \cite{Pelissetto02}). These behaviors follow from the fact that the singularity nearest to the origin in the complex $p^2$ plane is 9$m^2$ (4$m^2$) because the Minkowskian version of the theory has a three- (two-) particle cut in the symmetric (broken) phase respectively\footnote{The situation is in fact a little more complicated in the broken phase, see \cite{Pelissetto02}, but this has no qualitative impact on our arguments.}. Our argument relies on the fact that any regulator acts as a (momentum dependent) mass term. Thus, the critical theory regularized by  $R_k(q^2)$ should be similar to the non-critical (massive) theory and should therefore also have a convergent expansion in $p^2/k^2$ --which is nothing but the DE-- with a finite radius of convergence that we call $\cal R$ typically between 4 and 9 as we show below. 

At criticality and for  $s=6$, the analogue of Eq. (\ref{DE-massive-case}) is
\begin{eqnarray}
 \frac{\Gamma^{(2)}_k(p,\phi)+R_k(0)}{\Gamma^{(2)}_k(0,\phi)+R_k(0)}&&\hspace{1mm}=1+
    \frac{Z_k p^2 + W^a_k p^4 + X^a_kp^6}{U_k''+R_k(0)}\nonumber\\
     &&\hspace{-23mm}\xrightarrow[k\to 0]{} 1+\frac{p^2}{m_{\text{eff}}^2}+
    \frac{w_{a}^* {v^*}''}{{z^*}^2}\frac{p^4}{m_{\text{eff}}^4}+
    \frac{x_{a}^*{{v^*}''}^2}{{z^*}^3}\frac{p^6}{m_{\text{eff}}^6}
    \label{DE-criticality}
\end{eqnarray}
where the field dependence in the r.h.s. has been omitted, 
primes denote derivation wrt $\phi$ or $\tilde\phi$,
$u^*, z^*, w_a^*,x_a^*$ stand for the dimensionless functions of $\tilde\phi$ at the fixed point, and $m_{\text{eff}}^2=k^2{v^*}''/z^*$ with ${v^*}''={u^*}''+R_k(0)/Z^0_k k^2$.

If the two expansions (\ref{DE-massive-case}) and (\ref{DE-criticality}) are indeed similar then $m_{\text{eff}}^2$ 
must be the mass generated by the regulator and the coefficients of $p^4/m_{\text{eff}}^4$ and  $p^6/m_{\text{eff}}^6$ 
must be analogous to $c_2$ and $c_3$ in Eq. (\ref{DE-massive-case}). 
As for $m_{\text{eff}}$, if it is indeed generated by the regulator, it must be of order $R_k(q^2=0)\simeq\alpha k^2$ (see discussion below and Eq.~(\ref{regulators})).

It is  known that the $c_n$ in Eq. (\ref{DE-massive-case}) form an alternating series and that they are very small \cite{Pelissetto02}: 
In the symmetric phase, $c_2=-4\times10^{-4}$ and $c_3=0.9\times10^{-5}$ and in the broken phase, $c_2\simeq- 10^{-2}$ and $c_3\simeq4\times10^{-3}$. Together with the fact that the series in Eq. (\ref{DE-massive-case}) has a finite radius of convergence, this suggests that it not only converges but that it does so rapidly.

\begin{figure}
     \begin{picture}(216,165)
      \put(0,0){\includegraphics[width=216pt]{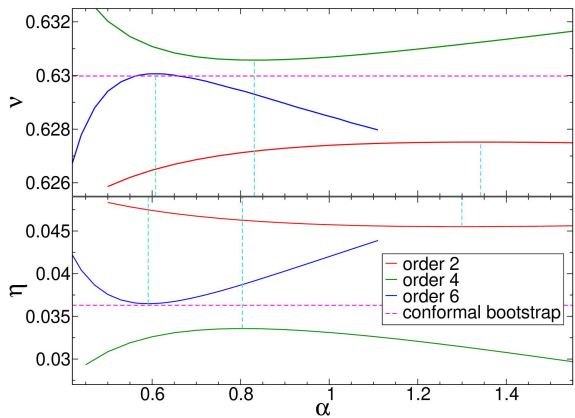}}
     \end{picture}
     \label{fig_opt}
     \caption{Exponent values $\nu(\alpha)$ and $\eta(\alpha)$ at different orders of the DE for regulator (\ref{regulator-exp}). 
     Vertical lines indicate $\alpha_{\text{opt}}$. LPA ($s=0$) results do not appear within the narrow ranges of values chosen here
      (see Table~\ref{simple-table}).
     }
     \label{pms}
\end{figure}

Let us now discuss  the role and shape of the regulator function $R_k(q^2)$. 
If no approximation of the exact flow equation (\ref{wetterich}) were made, all physical quantities would be independent 
of the choice of regulator. 
However the DE ---like any approximation scheme--- introduces an influence of the choice of $R_k$ on the end results
~\footnote{The choice of the normalization point $\tilde\phi_0$ also influences results, but it can be shown that its influence is exactly the same as that of the overall prefactor $\alpha$ in front of $R_k$\cite{TBP}.}.
There exist some constraints and a priori guidelines to choose $R_k$ so that its influence stays minimal.
First, $R_k$ must freeze the small momentum modes $\varphi(\vert {\bf q}\vert<k)$ in ${\cal Z}_k[J]$ (Eq.(\ref{partition})), 
so that they decouple from the long distance physics. It must also leave unchanged the large momentum modes $\varphi(\vert {\bf q}\vert>k)$. 
Second, the DE being a Taylor expansion of the $\Gamma^{(n)}_k(\{{\bf p}_i\})$ in powers of ${\bf p}_i\cdot {\bf p}_j/k^2$ (around 0), 
it is valid provided ${\bf p}_i\cdot {\bf p}_j/k^2< {\cal R}$.
This implies that whenever the $\Gamma^{(n)}_k$'s are replaced in a flow equation by their DE, 
the momentum region beyond ${\cal R}$  must be efficiently cut off. 
This is the role of the  $\partial_t R_k(q^2)$ term in Eq. (\ref{wetterich}). It suppresses this kinematic sector in the integral 
over the internal momentum ${\bf q}$ if $R_k(q^2)$ almost vanishes for $\vert {\bf q}\vert \gtrsim k$. 
On the other hand, modes $\varphi(\vert {\bf q}\vert<k)$ are almost frozen if $R_k(q^2)$ is of order $k^2$ for  $\vert {\bf q}\vert <k$. 
These two characteristic features give the overall shape of $R_k(q^2)$.
Note also that if a non-analytic regulator is chosen, one must make sure that the non-analyticities thus introduced in the complex plane of $q^2$  are further than ${\cal R}$ from the origin.
Finally, at order $s$ of the DE the flow equations of the functions involve $\partial_t R_k(q^2)$ and 
 $\partial^{n}_{q^2}R_k(q^2)$ from order $n=1$ to $s/2$. Since the DE is performed around $q=0$,
it is important that these derivatives decrease monotonically: if not,
a ``bump" at a finite value $q^2=q_0^2>0$ would magnify a region around $q_0$ which is less accurately described 
by the DE~\footnote{We have numerically checked this by comparing two very similar regulators except that one has a bump in $\partial_t R_k(q^2)$ and its derivatives with respect to $q^2$ around $q^2=k^2$  \cite{TBP}.}.

Taking into account all the prerequisites above, we have considered either regulators that are $C^\infty$ 
in the complex plane of $q^2$, decay rapidly but do not vanish for $q>k$,
or functions that vanish identically for $q>k$, are not $C^\infty$ but are sufficiently differentiable for regularizing the DE 
at the order $s$ studied, and have their derivatives as small as possible for $q\simeq k$. 
Specifically,  we used:
\begin{subequations}
\begin{align}
W_k(q^2) &= \alpha Z^0_k k^2 \, y/(\exp(y) - 1) \label{regulator-wetterich}\\ 
\Theta^n_k(q^2) &= \alpha Z^0_k k^2 \, (1-y)^n \theta(1-y)   \;\;\;n\in\mathbb N  \label{regulator-theta}\\
E_{k}(q^2) &= \alpha Z^0_k k^2 \,\exp(-y)  \label{regulator-exp}
\end{align}
\label{regulators}
\end{subequations}
where $y=q^2/k^2$. We show in the Supplemental Material that $\Theta^\infty_k(q^2)$ is equivalent to $E_{k}(q^2)$.

\begin{table}[tp]
\caption{3D Ising critical exponents obtained with regulator (\ref{regulator-exp}) at orders $s=0$ (LPA) to 6. 
Absolute distances between these values and the near-exact
conformal bootstrap \cite{Kos:2014bka} ones are given by $|\delta\nu|$ and $|\delta\eta|$.
Monte-Carlo \cite{Hasenbusch10}, High-temperature expansion \cite{Campostrini:2002cf}, 
and 6-loop perturbative RG values \cite{Guida:1998bx} are also given for comparison.}
\label{simple-table} 
\begin{ruledtabular}
\begin{tabular}{lllll}
D.E. &  $\nu$        & $|\delta\nu|$     &  $\eta$ & $|\delta\eta|$	\\
$s=0$   & 0.65103  	 &  0.02106          &  0               & 0.03630	\\
$s=2$   & 0.62752    &  0.00245          &  0.04551         & 0.00921 \\
$s=4$   & 0.63057    &  0.00060          &  0.03357         & 0.00273\\
$s=6$   & 0.63007    &  0.00010          &  0.03648         & 0.00018\\
\hline
conf. boot.  & 0.629971(4)  &	    & 0.0362978(20) &\\
6-loop       & 0.6304(13)   &       & 0.0335(25)    & \\
High-T.      & 0.63012(16)  &       & 0.03639(15)   & \\
M.-C.        & 0.63002(10)  &       & 0.03627(10)   & 
\end{tabular}
\end{ruledtabular}
\end{table}

We now present our results~\footnote{Previous works at LPA and order 2 
\cite{Wegner:1972ih,Hasenfratz:1985dm,Tetradis94,Morris:1994ie,Seide:1998ir,Litim:2002cf,Canet03a,Berges:2000ew,Delamotte:2007pf} 
and order 4 \cite{Canet03b} exist, but often considered slightly different flow equations and sometimes resorted to 
truncated Taylor expansions of the fields, see Section B of the Supplemental Material. A detailed discussion will be given in \cite{TBP}.}.
We focus here on the three-dimensional (3D) case, for which
near-exact results are provided by conformal bootstrap \cite{ElShowk:2012ht,El-Showk:2014dwa,Kos:2014bka}, 
but we have obtained similar in two dimensions \cite{TBP}.
For each regulator function $R_k$, we have calculated the critical exponents $\nu$
 (associated with the divergence of the correlation length) and $\eta$, 
 as well as the different ratios appearing in Eq. (\ref{DE-criticality}), at orders $s=0$ (LPA), 2, 4, and 6. 
 (Numerical details can be found in Section D of the Supplemental Material.)
These quantities depend on the parameters of $R_k$, that is, for the regulators (\ref{regulators}), on
$\alpha$ that we typically  vary  in the range $[0.1,10]$. 

Each regulator function we studied yielded very similar results. We first discuss those obtained with~(\ref{regulator-exp}). 
In Fig.~\ref{pms}, we show the curves $\nu(\alpha)$ and $\eta(\alpha)$ for orders $s=2$ to 6.
 At each order, exponent values exhibit a maximum or a minimum at some value $\alpha_{\rm opt}$ as $\alpha$ varies. 
 Following a ``principle of minimal sensitivity" \cite{Stevenson:1981vj,Canet03a}, 
we select the values $\nu(\alpha_{\rm opt})$ and $\eta(\alpha_{\rm opt})$ taken at the extrema as 
our best estimates. Note that this is the situation closest to the exact theory, for which there is no dependence on the regulator.
At a given order $s$, $\alpha_{\rm opt}^{(\nu)}$ and $\alpha_{\rm opt}^{(\eta)}$ are close but different, and their difference decreases fast with increasing $s$,  see Fig. \ref{pms}.

Important remarks are in order.
For each exponent, increasing the order $s$: 
(i) extrema alternate between being given by a maximum and a minimum;
(ii) the local curvature at $\alpha_{\rm opt}$ increases;
(iii) strikingly, the exponent values at $\alpha_{\rm opt}$ essentially alternate around and converge very fast to values very close to 
the conformal bootstrap ``exact" ones. (At order 6, the optimal value of $\nu$ `crosses' the exact value, but these 2 numbers 
coincide up to 3 or 4 significant digits, see Fig.~\ref{pms} and Table~\ref{simple-table}.) 
The increase of curvature at $\alpha_{\rm opt}$ and the accompanying faster variations of exponent values with $\alpha$ as $s$ is increased
imply that it is crucial to work with the optimal values given by the extrema, that is $\nu(\alpha_{\rm opt}^{(\nu)})$ and $\eta(\alpha_{\rm opt}^{(\eta)})$.
This fast, alternating convergence is due to the alternating nature of the series of coefficients $c_n$. 
The speed of convergence is also in agreement with our considerations above about the radius of convergence $\cal{R}$ of the DE at criticality:
the amplitude of the oscillations of the optimal values considered as functions of $s$ decreases 
typically by a factor between 4 to 10 at each order $s$ (Table~\ref{simple-table}).

As mentioned above, all regulators we studied yield very similar results. 
For each exponent, the dispersion of values (over all regulators studied) 
typically also decreases by a factor 4 to 8 when going from one order $s$ to the next,
something we interpret as another manifestation of the radius of convergence of the DE, see Table \ref{family-values}. 
We also noticed that regulators not satisfying our prerequisites very well typically yield ``worse" results,
somewhat away from those given by the set of good regulators (\ref{regulators}) \cite{TBP}.
Our extensive exploration of regulators, including some multi-parameter ones not described here, 
thus leads us to conjecture the existence, for a given  exponent and a given order of the DE, of an optimum-optimorum value,
a ``ceiling" ---or a ``floor", depending on the exponent and the order considered--- that cannot be passed by any regulator 
(taken at its optimal parameter value $\alpha_{\rm opt}$). 
In particular, at LPA level, we did not find any regulator able to yield a $\nu$ value
below the one given by the Wilson-Polchinski approach $\nu_{\rm WP}=0.6496$ \cite{Hasenfratz:1985dm}. 
We recall that this value is the one given by 
regulator $\Theta^1_k$, which thus appears, under our conjecture, as {\it the} optimal regulator at LPA level 
\cite{Litim:2002cf,pawlowski2017}.

\begin{table}[tp]
\caption{Exponent values given by the middle of the range of values observed 
over the family of regulators (\ref{regulator-wetterich}--\ref{regulator-exp}).
Error bars are simply given by the half range. Extrapolation to asymptotic ($s\to\infty$) values are obtained by fitting the finite-$s$ ones (see text).}
\begin{ruledtabular}
\begin{tabular}{lll}
derivative expansion &  $\nu$               &  $\eta$	\\
$s=0$ (LPA)          & 0.651(1)  	        &  0  	\\
$s=2$      	         & 0.6278(3) 	        &  0.0449 (6) 	\\
$s=4$                & 0.63039(18)          &  0.0343(7)     \\
$s=6$                & 0.63012(5)           &  0.0361 (3)      \\
$s\to\infty$         & 0.6300(2)            & 0.0358(6) \\
\hline
conformal bootstrap  & 0.629971(4) 	    & 0.0362978(20) 
\end{tabular}
\end{ruledtabular}
\label{family-values} 
\end{table}

The above conjecture, if adopted, allows to order regulators by increasing quality. Pending a proof, or, better, a constructive
method to determine optimal regulators, we propose to use, at each order $s$, 
the range of exponent values over a family of ``reasonable" regulators 
to define typical values (given by the middle of this range) and error bars (given by the half-range,
which may appears as a conservative estimate). The resulting numbers are in Table~\ref{family-values}.
Similarly, to estimate asymptotic ($s\to\infty$) values for a given problem treated by the DE
we propose to extrapolate results obtained at low orders taking into account the facts uncovered above:
The exponents, considered as functions of $s$, should have a monotonous as well as an oscillating contribution. 
For instance $\nu(s)= \nu_\infty + a_\nu \beta^{-s/2}+ b_\nu (-1)^{s/2} \beta^{-s/2}$ where, typically, $4\le\beta\le9$ (given by the radius of convergence)
and $a$ and $b$ are unknown coefficients. 
By considering all the regulators we have studied as well as all values of $\beta$ between 4 and 9 we obtain a dispersion of asymptotic estimates,
whose mean and maximal extent give us the following final numbers and the associated error bars 
(Table~\ref{family-values}, see also Section E of the Supplemental Material):
$\nu=0.6300(2)$ and $\eta=0.0358(6)$. Remarkably, these are in excellent agreement with conformal bootstrap values 
$\nu=0.629971(4)$ and $\eta=0.0362978(20)$, and better than perturbative 6-loop ones.

We now come back to the momentum expansion of $\Gamma^{(2)}_k(p,\phi)+R_k(0)$ in the light of our results.
We emphasized above that if $m_{\text{eff}}$ in Eq.~(\ref{DE-criticality}) is the mass generated by the regulator  
then it must be of order $R_k(q^2=0)\simeq\alpha k^2$, which implies that we should have ${u^*}''/z\propto \alpha$. 
Remarkably, this relation is satisfied to a high accuracy for all regulators we have studied 
when using our optimal values $\alpha_{\rm opt}$, see Section E of the Supplemental Material.
We have also checked for all regulators and all $\tilde\phi$ that $w_{a}^* {v^*}''/{z^*}^2<0$ and  $x_{a}^*{v^*}''^2/{z^*}^3>0$ 
in agreement with the signs of $c_2$ and $c_3$. 
The ratio $r= x_a^*{u^*}''/(w_a^* z^*)$, which plays a role analogous to $c_3/c_2$,  
varies between $-\tfrac{1}{20}$ for $\tilde\phi$ around $\tilde\phi_{\text{min}}$, the minimum of the potential,  
and $-\tfrac{1}{4}$ at large $\tilde\phi$ and is largely regulator independent (see Section F of the Supplemental Material). 
These values correspond typically to what is found in the symmetric and broken phases respectively, 
which is expected for a regularized theory at criticality. 

We now go a step further and explain the behavior of the coefficients of $p^4$ and $p^6$ in Eq. (\ref{DE-criticality}).
We know that at criticality, when $ p\gg k$, $\Gamma_k^{(2)}(p,0)\simeq \Gamma_{k=0}^{(2)}(p,0)\propto p^{2-\eta}$. 
On the other hand, when $p\ll k$, $\Gamma_k^{(2)}(p)$ is given by Eq. (\ref{DE-criticality})  at $\phi=0$. 
Matching these two expressions for $p\sim k$, we find a simple analytic representation of the form 
$\Gamma_k^{(2)}(p)\simeq A p^2(p^2+b k^2)^{-\eta/2} + m_k^2$ where $A$ and $b$ are constants and $m_{k=0}=0$.
Expanded in powers of $p^2/k^2$, this expression yields an alternating series with a negative coefficient starting from $p^4$ 
and a positive one for $p^6$ as in Eq.~(\ref{DE-massive-case}). Moreover, all coefficients of the series from $p^4$ are proportional to $\eta$, 
which makes all of them naturally small, again as in Eq. (\ref{DE-massive-case}). 
We therefore conclude that the DE is a convergent expansion with (i) a finite radius of convergence typically between 4 and 9 and, 
(ii) a rapid convergence because all the  coefficients of the $(p^2/k^2)^n$  terms with $n\ge2$ are proportional to $\eta$, 
which is small for the 3D Ising model \footnote{In $2D$, $\eta$ is about 7 times larger, which in turn explains why the DE does not converge 
as fast then \cite{TBP}.}.

In summary, we have shown that the derivative expansion often used in NPRG studies has a finite radius of convergence and we provided
guidelines for choosing the regulator function at the heart of the procedure.
Using the Ising model in three dimensions as a testing ground, we find fast convergence of critical exponents to their exact values, 
irrespective of the well-behaved regulator used, in full agreement with our general arguments. Our findings naturally extend to
many other models --those having a unitary Minkowskian extension-- 
and to other NPRG approximations such as the Blaizot-Mend\'ez-Wschebor scheme~\cite{Blaizot:2005xy,Benitez:2009xg,Benitez:2011xx}.
This establishes firmly that the NPRG approach is not only versatile, being able to deal with any equilibrium or non-equilibrium model, but also quantitative, providing accurate results even at low orders.

This work was supported by Grant 412FQ293 of the CSIC (UdelaR) Commission and Programa
de Desarrollo de las Ciencias B\'asicas (PEDECIBA),
Uruguay and ECOS Sud U17E01.
IB acknowledges the support of the Croatian Science Foundation Project IP-2016-6-7258 
and the QuantiXLie Centre of Excellence, a project cofinanced by the Croatian Government and European Union through the European Regional Development Fund - the Competitiveness and Cohesion Operational Programme (Grant KK.01.1.1.01.0004). IB also thanks the LPTMC for its hospitality and the CNRS for funding during the spring of 2019. BD thanks L. Canet,  N. Dupuis, C. Fleming, J. Pawlowski and M. Tissier for discussions and remarks about an early version of the manuscript.

\bibliography{biblio}
\bibliographystyle{apsrev4-1}

\pagebreak


\begin{center}
\Large \bf
  SUPPLEMENTAL MATERIAL
\end{center}

\subsection{Ansatz for $\Gamma_k$ at order 6}

At order 6 of the derivative expansion, the ansatz for $\Gamma_k[\phi]$ reads: 
\begin{eqnarray}
&\hspace{-0.5cm}  \Gamma_k[\phi]&= \int d^dx \Big[U_k(\phi) + \tfrac{1}{2} Z_k(\phi) (\partial\phi)^2 \nonumber\\
&&\!\!\!\!\!\!\! + \tfrac{1}{2} W^a_k(\phi)(\partial_\mu\partial_\nu\phi)^2 + \tfrac{1}{2} \phi W^b_k (\phi)(\partial^2\phi)(\partial\phi)^2 \nonumber\\
&&\!\!\!\!\!\!\! + \tfrac{1}{2}  W^c_k (\phi)\left((\partial\phi)^2\right))^2+ \tfrac{1}{2} \tilde X^a_k (\phi)(\partial_\mu\partial_\nu\partial_\rho\phi)^2 \nonumber\\
&&\!\!\!\!\!\!\! + \tfrac{1}{2} \phi \tilde X^b_k (\phi)(\partial_\mu\partial_\nu\phi)(\partial_\nu\partial_\rho\phi)(\partial_\mu\partial_\rho\phi) \nonumber\\
&&\!\!\!\!\!\!\! + \tfrac{1}{2}  \phi\tilde X^c_k(\phi)\left(\partial^2\phi\right)^3+\tfrac{1}{2}  \tilde X^d_k (\phi)\left(\partial^2\phi\right))^2 (\partial\phi)^2\nonumber \\
&&\!\!\!\!\!\!\! + \tfrac{1}{2} \tilde X^e_k (\phi)(\partial\phi)^2 (\partial_\mu\phi)(\partial^2\partial_\mu\phi)+ \tfrac{1}{2} \tilde X^f_k (\phi)(\partial\phi)^2 (\partial_\mu\partial_\nu\phi)^2 \nonumber\\
&&\!\!\!\!\!\!\! + \tfrac{1}{2}  \phi\tilde X^g_k (\phi)\left(\partial^2\phi\right)\left((\partial\phi)^2\right)^2+ \tfrac{1}{96} \tilde{X}^h_k (\phi) \left((\partial\phi)^2\right)^3 \Big].
\label{ansatz-order6-complet}
\end{eqnarray}
with
\begin{eqnarray}
&&\tilde{X}_a=X_a, \; \tilde{X}_b=X_b-2 X_c, \; \tilde{X}_c=X_c\nonumber\\
&&\tilde{X}_d=\tfrac{1}{6}X_d - \tfrac{2}{3}(X_c + \phi X_c')\nonumber\\
&&\tilde{X}_e=\tfrac{1}{4}(X_d+X_e+X_f)-(X_c + \phi X_c')\nonumber\\
&&\tilde{X}_f=\tfrac{1}{12}X_d+\tfrac{1}{4}(X_e-X_f)-\tfrac{1}{3}(X_c + \phi X_c')\nonumber\\
&&\tilde{X}_g= 6X_g+4(2X_c'+ \phi X_c'') - X_d'-9X_e'-9X_f'\nonumber\\
&&\tilde{X}_h=X_h+20(3X_c''+ \phi X_c''')-5X_d''-7X_e''\nonumber\\
&& {\hspace{0.9cm}}-9X_f''+2X_g'
\end{eqnarray}
The vertex functions $\Gamma^{(n)}_k(\{p_i\},\phi)$ where $\phi$ is a constant field can be computed in Fourier space  
from (\ref{ansatz-order6-complet}) in terms of  $U_k(\phi),\cdots,\tilde{X}_k^h(\phi)$. 
The definition of these functions follows from this calculation and yields for instance: 
$Z_k(\phi)=\partial_{p^2}\Gamma_k^{(2)}(p;\phi)\vert_{{\bf p}=0}$. In general, the $\tilde{X}_k(\phi)$ functions are given by  linear combinations of $\Gamma^{(n)}_k(\{{\bf p}_i\},\phi)$ functions computed in different configurations of their momenta. To avoid this difficulty, it is convenient to redefine some of the functions $\tilde{X}_k(\phi)$ so that the new functions are given directly by one $\Gamma^{(n)}_k(\{{\bf p}_i\},\phi)$ in a simple configuration of momenta. This is the reason why we have worked with the ${X}_k(\phi)$  instead of the $\tilde{X}_k(\phi)$ functions. For instance:
\begin{eqnarray}
&&\,{\hspace{-0.8cm}} X^d_k (\phi)=\partial_{p_1^2 p_2^2 p_3^2}
\Gamma^{(4)}_k({\bf p}_1,{\bf p}_2,{\bf p}_3;\phi)_{\displaystyle\vert_{{\bf p}_i=0}}\nonumber\\
&&\,{\hspace{-0.8cm}} X^h_k (\phi)=\partial_{({\bf p}_1.{\bf p}_2)({\bf p}_3.{\bf p}_4)({\bf p}_5.{\bf p}_1)}
\Gamma^{(6)}_k({\bf p}_1,\dots,{\bf p}_5;\phi)_{\displaystyle\vert_{{\bf p}_i=0}}
\label{def-X}
\end{eqnarray}

\subsection{Flow equations, truncation issues}

Let us consider the function $Z_k(\phi)$ defined by
\begin{equation}
Z_k(\phi)= \partial_{p^2}\Gamma^{(2)}_k(p;\phi)\vert_{{\bf p}=0}
\label{def:Z}
\end{equation}
Its flow is given by the $p^2$ term of the flow of $\Gamma^{(2)}_k(p;\phi)$:
\begin{eqnarray}
\partial_t&&\hspace{-0.cm}\Gamma^{(2)}_k(p;\phi)=\int_q \dot{R}_k(q^2) G_k^2(q)\Big[-
\frac{1}{2}\Gamma^{(4)}_k({\bf p},{-\bf p},{\bf q},-{\bf q})+\nonumber\\
&&\hspace{-0.35cm}
\Gamma^{(3)}_k({\bf p},{\bf q},-{\bf p}-{\bf q})G_k({\bf p}+{\bf q})
\Gamma^{(3)}_k(-{\bf p},-{\bf q},{\bf p}+{\bf q})\Big].
\label{flowZ}
\end{eqnarray}
Let us now consider the flow of $Z_k(\phi)$ at order $s=2$ of the DE where:
\begin{equation}
\Gamma_k[\phi]= \int_x \left[U_k(\phi) + \frac{1}{2} Z_k(\phi) (\nabla\phi)^2 + O(\nabla^4) \right].
\label{s=2}
\end{equation}
The value of $\Gamma^{(3)}_k$ computed from (\ref{s=2}) is inserted in the term
$\Gamma^{(3)}_k({\bf p},{\bf q},-{\bf p}-{\bf q})\Gamma^{(3)}_k(-{\bf p},-{\bf q},{\bf p}+{\bf q})$
of Eq.(\ref{flowZ}).
Only terms at order 2 in the internal ($q$) and external ($p$) momenta are kept. Similarly, at order $s$, only terms at order $s$ are kept on the rhs of flow equations.

Note that this is {\it not} what most previous studies did, since usually all terms on the rhs are kept, in contradiction with
the spirit of a Taylor expansion. At order 6, cutting off higher-order terms on the rhs drastically simplifies expressions that would be otherwise very hard to handle numerically.
The differences in flow equations result in our exponent values being slightly different from those given previous works 
(references [7-8, 30-31, 37-42] of the main text). 
The differences may become more significant when, in addition, the functions  are Taylor expanded in powers of the field.

\subsection{Relation between the $\Theta_k^n$ and exponential regulators}

We show here that the $E_k(q^2)$ regulator is the $n\to\infty$ limit of the $\Theta_k^n(q^2)$ function.  
Recall first that two regulators that only differ by the definition of their scale $k$ lead to identical physical results when $k\to 0$. 
The regulator 
\begin{equation}
 \Theta_k^n(\alpha,q^2)=
 \alpha {Z^0_k} k^2\left(1-\frac{q^2}{k^2}\right)^n \theta\left(1-\frac{q^2}{k^2}\right) \;,   
\end{equation}
when we change its scale $k$ into $k'$ given by $k=\sqrt n\, k'$, becomes
\begin{equation}
 \Theta_k^n(\alpha,q^2)= \alpha Z^0_{k'} n {k'}^2\left(1-\frac{q^2}{n{k'}^2}\right)^n \theta\left(1-\frac{q^2}{n{k'}^2}\right)    
\end{equation}
that we now consider as a new, although equivalent, regulator depending on the scale $k'$. 
When $n\to\infty$ we have
\begin{equation}
\!\! \left(1\!-\!\frac{q^2}{n{k'}^2}\right)^n \!\! \to \exp \left(\!-\frac{q^2}{{k'}^2}\right) \;  {\rm and} \; \theta\left(1\!-\!\frac{q^2}{n{k'}^2}\right)\to 1.
\end{equation}
Thus, at very large values of $n$, $\Theta_k^n(q^2)$ becomes equivalent to the regulator 
$\alpha n Z_{0,k'} {k'}^2\exp(-q^2/{k'}^2)$. This imposes that when $n$ and $n'$ are large, the two regulators $\Theta_k^n(\alpha,q^2)$
and $\Theta_k^{n'}(\alpha',q^2)$ are (almost)
equivalent if  $ \alpha\, n=\alpha' n'$ because they are equivalent to the exponential regulator $E_k(q^2)$. We have numerically checked that, for $n,n'>10$, this relation is satisfied to a very good accuracy when using the optimal values $\alpha_{\rm opt}(n)$ and $\alpha_{\rm opt}(n')$.

\subsection{Numerical implementation}

We directly solved for the fixed point solutions of the coupled integro-differential flow equations using a Newton-Raphson method. 
To obtain good enough initial guesses, we first simulated directly the equations (using explicit Euler time-stepping) 
from some bare initial condition, proceeding by ``dichotomy"
to get near the fixed point at large RG times. 
All our results have been checked against changing resolution, the extent of the field domain considered,
the accuracy with which integrals are calculated, and the order at which derivatives are estimated. Typical choices are:
discretization of the field $\rho=\phi^2$ over a regular grid, field domain extending from 0 to 3 times the minimum of the potential with free boundary conditions, derivatives calculated over 7 grid points, integrals computed with the Gauss-Kronod method and a relative accuracy of order $10^{-12}$.

\begin{figure}[t!]
     \begin{picture}(216,197)
      \put(0,0){\includegraphics[width=216pt]{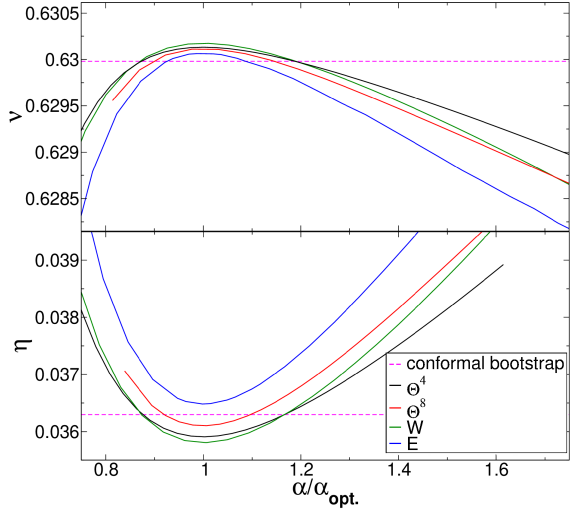}}
     \end{picture}
     \caption{Exponent curves $\nu(\alpha)$ and $\eta(\alpha)$ at order $s=6$ for different regulators. Conformal bootstrap values
     are indicated by the dashed horizontal lines.}
     \label{fig_suppl_II}
\end{figure}

\begin{table}[h!]
\caption{\label{table_suppl} Ising critical exponents in $d=3$ obtained with various regulators. The numbers in parentheses give the distance of the results to the conformal bootstrap values taken here as the exact ones.}
\begin{ruledtabular}
\begin{tabular}{llll}
      & \hspace{-0.5cm}regulator          &  $\nu$         &  $\eta$ \\
\hline
LPA   &$W$             & 0.65059(2062)          &  0            \\
      &$\Theta^1$      & 0.64956(1959)          & 0               \\
      &$\Theta^3$      & 0.65003(2006)          &  0                \\
      &$\Theta^4$      & 0.65020(2023)          &  0            \\
      &$\Theta^8$      & 0.65056(2059)          &  0            \\
      &$E$             & 0.65103(2106)          &  0      \\           
\hline   
$O(\partial^2)$
     &$W$              & 0.62779(218)          &  0.04500(870)  \\
     &$\Theta^2$       & 0.62814(183)          &  0.04428(798)  \\
     &$\Theta^3$       & 0.62802(195)          &  0.04454 (824)  \\
     &$\Theta^4$       & 0.62793(204)          &  0.04474(844)  \\
     &$\Theta^8$       & 0.62775(222)          &  0.04509(879)  \\
     &$E$              & 0.62752(245)          &  0.04551(921)  \\
\hline
$O(\partial^4)$
      &$W$           & 0.63027(30)           & 0.03454(176)  \\
      &$\Theta^3$    & 0.63014(17)           & 0.03507(123)    \\
      &$\Theta^4$    & 0.63021(24)           & 0.03480(150)  \\
      &$\Theta^8$    & 0.63036(39)           & 0.03426(204)   \\
      &$E$           & 0.63057(60)           & 0.03357(272)  \\
\hline
$O(\partial^6)$
      &$W$           & 0.63017(20)          & 0.03581(49)  \\
      &$\Theta^4$    & 0.63013(16)          & 0.03591(39)  \\
      &$\Theta^8$    & 0.63012(15)          & 0.03610(20)  \\
      &$E$           & 0.63007(10)          & 0.03648(18)  \\
      \hline
conf. boot. &        & 0.629971(4) 	    & 0.0362978(20) 
\end{tabular}
\end{ruledtabular}
\end{table}

\subsection{Exponent values}

At each order $s$, all regulators we used provide similar-looking $\nu(\alpha)$ and $\eta(\alpha)$ curves. 
Figure~\ref{fig_suppl_II} shows them, centered around the $\alpha_{\rm opt}$ position, at order 6. Note that at this order 
the dispersion of exponent values at  $\alpha_{\rm opt}$ is of the order of the distance to the exact values. 
The range of exponent values found at $\alpha_{\rm opt}$ was used to define our `final' estimates (taken to be the value at the middle of the range), 
and their error bars (given by the half-range) in Table~II of the main text.

Exponent values for the $\Theta^n$ regulators were found to vary monotonously with $n$, from the minimal value $n=1+s/2$ possible at order $s$
to infinity, represented by the exponential regulator $E$. In addition to these, we also used the $W$ regulator. Table ~\ref{table_suppl} gives  
typical values found, together with the quasi-exact ones provided by conformal bootstrap.
For simplicity, we only show, at each order, the digits that vary when changing the regulator.

These results show the reduction in the regulator dependent scatter, as the order of the derivative expansion is increased, despite the fact that the optimization curves become steeper with increasing orders, as seen in Fig. 1 of the main text. 

\begin{figure}[t!]
     \begin{picture}(216,108)
      \put(0,0){\includegraphics[width=216pt]{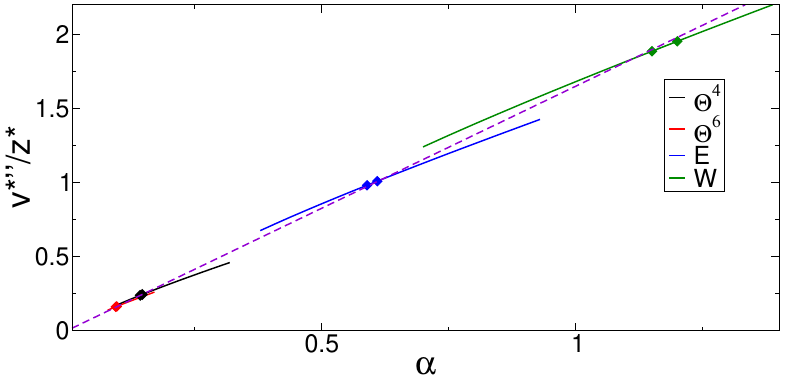}}
     \end{picture}
     \caption{Squared dimensionless mass generated by the regulator $\tilde{m}^2_{\rm eff}(\phi_{\rm min})= {v^*}''/z^*|_{\phi_{min}}$ as a function of the regulator parameter $\alpha$ for several different regulators. Diamonds show the optimal values $\alpha_{\rm opt}$ for $\eta$ (for a smaller value of $\alpha$) and $\nu$ (for a larger value of $\alpha$). The dashed line has slope 1.65.
These data have been obtained at $\phi_{\rm min}$ the minimum of the dimensionless fixed point potential, but choosing another point yields similar results.}
     \label{fig_suppl}
\end{figure}

\begin{figure}[b]
     \begin{picture}(216,120)
      \put(0,0){\includegraphics[width=216pt]{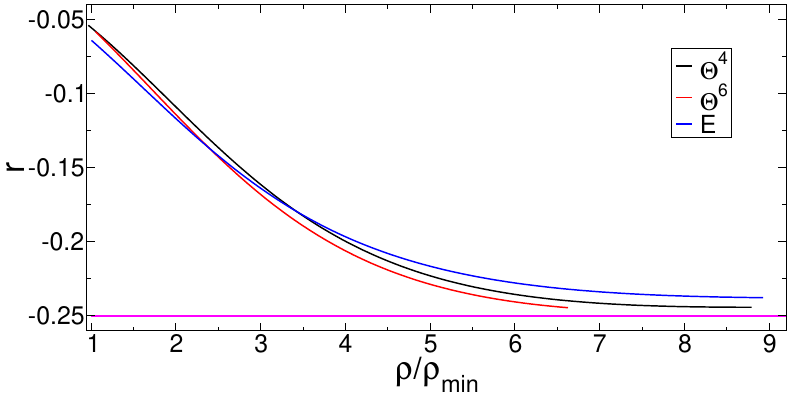}}
     \end{picture}
     \caption{The ratio $r= x_a^*{u^*}''/(w_a^* z^*)$ as a function of $\tilde{\rho}=\tilde{\phi}^2/2$. The line $r=0.25$ is a guide for the eyes.}
     \label{fig_c3c2}
\end{figure}

\subsection{Numerical results about the momentum expansion of $\Gamma^{(2)}_k(p,\phi)+R_k(0)$} 

The dependence of the dimensionless squared mass $\tilde{m}^2_{\rm eff}(\phi)$ on the regulator parameter $\alpha$ is predominantly linear, for all regulators considered (Fig. \ref{fig_suppl}). For every regulator, varying $\alpha$, we find a small deviation from the linear dependence, that can be well fitted by a logarithm.
Strikingly, at the optimal values $\alpha_{\rm opt}$, $\tilde{m}^2_{\rm eff}(\phi)$ is exactly on a straight line passing through the origin.

In Figure~\ref{fig_c3c2}, we show how the ratio $r= x_a^*{u^*}''/(w_a^* z^*)$, which plays a role analogous to $c_3/c_2$, varies with
$\tilde{\phi}$. The ratio $r$ is always negative as expected. For values of $\tilde \rho$ close to the minimum of the potential $\rho_{\text{min}}$, the ratio is order $-1/20$. On the other hand, when $\tilde\rho$ goes to infinity $r$ approaches $-1/4$. This shows that $r$ varies between the typical values of $c_3/c_2$ corresponding to the symmetric phase for $\rho\sim\rho_{\text{min}}$ and the value $-1/4$ of the broken phase. This is consistent with the fact that for large values of the external field the symmetry is broken and the theory has a gap with a threshold for multiparticle states  at $p^2=-4 m_{\rm eff}^2$.

\end{document}